\documentclass[apj]{emulateapj}   	
\pdfoutput=1
\usepackage{geometry}                		
\usepackage{amssymb}
\usepackage[dvipsnames]{xcolor}

\usepackage{multirow}
\usepackage{amsmath}
\usepackage{aas_macros}

\begin{document}
\title{Filtering interlopers from galaxy surveys}
\author{
Kaze Wong
}
\affil{The Chinese University of Hong Kong, Shatin, NT, Hong Kong SAR, The People's Republic of China\\
McWilliams Center for Cosmology, Department of Physics, Carnegie Mellon University, 5000 Forbes Ave, Pittsburgh, PA, 15213, U.S.A.}

\author{
Anthony R. Pullen
}
\affil{McWilliams Center for Cosmology, Department of Physics, Carnegie Mellon University, 5000 Forbes Ave, Pittsburgh, PA, 15213, U.S.A.}
\author{
Shirley Ho
}
\affil{McWilliams Center for Cosmology, Department of Physics, Carnegie Mellon University, 5000 Forbes Ave, Pittsburgh, PA, 15213, U.S.A.}
\renewcommand\abstractname{\textbf{\Large Abstract}}
\begin{abstract}
We present Intercut, a Python-based program that applies secondary line identification and photometric cuts to mock galaxy surveys, in order to simulate interloper identification. This program can be used to optimize the removal of interloper contamination in upcoming surveys. Intercut reads a mock galaxy survey and an emission line sensitivity and simulates interloper removal through secondary line identification and broad-band photometry. This program is designed to use the COSMOS mock catalog, although the program can be modified for an alternative mock catalog. The output of the program returns an interloper fraction for each emission line, as well as the total fraction over all lines, as a function of redshift. We test Intercut by predicting interloper rates for the WFIRST emission line sensitivity, finding agreement with previous work. This program is publically available on Github\footnote{ https://github.com/kazewong/Intercut}.   
\end{abstract}

\section{INTRODUCTION}
A new era of large galaxy surveys is on the horizon. Upcoming surveys like Euclid \citep{2011arXiv1110.3193L} and the Wide-Field InfraRed Survey Telescope (WFIRST) \citep{2013arXiv1305.5422S}, each of which comprise over a billion galaxies, will obtain new constraints on dark energy by mapping the large-scale structure (LSS) of our universe up to redshift $z\sim 2$ (ESA/SRE 2011). Since the redshift range of these surveys is extended to a higher limit than previous ones, the signal to noise ratio (SNR) and the number of emission lines in each galaxy's spectrum is expected to be lower compared to the Sloan Digital Sky Survey (SDSS) \citep{2000AJ....120.1579Y}.

Since fewer lines are observed, it will be more difficult to confirm the identity of a particular emission line. This can lead to misidentifying interloping emission lines as the survey's intended one, producing an incorrect estimation of the galaxy's redshift. These interlopers will bias the measured matter power spectrum , along with the cosmological parameters. For example, it was shown in \cite{2015arXiv150705092P} that an interloper rate of $0.15\%$ would significantly bias the WFIRST measurement of the growth rate, a gravity probe within LSS. 

Two simple methods can help distinguish interlopers from emission line galaxies, namely secondary line identification and photometric cuts \citep{2007ApJ...660...62K}. Although more complex methods for removing interlopers\citep{2008ApJ...684...88N,2013arXiv1303.4722M} exist, they are limited by survey parameters, while the simple methods can be applied more generally and to each individual galaxy. Recently, \cite{2015arXiv150705092P} used these methods to predict an interloper fraction of $0.2\%$ for the WFIRST H$\alpha$ survey, and up to $10\%$ for the WFIRST OIII survey. 

In this paper, we introduce Intercut, a program that applies secondary line identification and photometric cuts to a mock galaxy survey, in order to predict the potential interloper fraction for any upcoming survey. The mock catalog used in the test run is the COSMOS Mock Catalog (CMC) \citep{2009ApJ...690.1236I, 2011A&A...532A..25J}. Aside from the catalog, an emission line sensitivity file containing information on the noise of the spectrograph as a function of wavelength, the noise for each photometric band, and the photometric cuts used is needed. 

The Intercut workflow is as follows: First, Intercut reads data from the catalog's FITS file, then a calibration of the galaxy fluxes using a user-defined luminosity function will be executed if needed. Next, emission line and photometric band fluxes are perturbed. Finally, secondary line identification and photometric cuts are applied, and the resulting interloper fractions are computed. We test Intercut by computing interloper fractions assuming WFIRST's sensitivity, finding it agrees with previous work. In addition to the individual interloper fractions for the each line, the total interloper fraction over all lines is calculated.

The paper is structured as follows: In section 2, we gives a brief review of the interlopers' effects on the matter power spectrum and cosmological parameters. In section 3, three major functions of the program are presented: calibrating luminosity functions, perturbing emission lines and photometric band fluxes, and identifying interlopers. In section 4,we present sample output for WFIRST.

\section{INTERLOPER REVIEW}
We define the  interloper fraction as  in \cite{2015arXiv150705092P},

\begin{equation}
f({\lambda}_{SEL}-{\lambda}_{Int},{z}_{
})=\frac{{N}_{Int}({z}_{Int})}{{N}_{SELG}({z}_{SELG})+{N}_{Int}({z}_{Int})},
\end{equation}
In Eq.(1), ${N}_{SELG}$ represents the number of survey emission line galaxies at redshift ${z}_{SELG}$, and ${N}_{Int}$  represents the number of interlopers at redshift ${z}_{Int}$. By equating the observed wavelengths of the survey and interloper emission lines, ${z}_{Int}$ in fact depends on ${z}_{SELG}$:
\begin{eqnarray}
{\lambda}_{obs} &=&{\lambda}_{rest}(z+1)\\
{\lambda}_{Int}({z}_{Int}+1) &=&  {\lambda}_{SELG}({z}_{SELG}+1)\nonumber\\
{z}_{Int} &=&  \frac{{\lambda}_{SELG}({z}_{SELG}+1)}{{\lambda}_{Int}} -1 \label{eqn},
\end{eqnarray}
Surveys can also have multiple interloping lines; the expression defined in Eq. (1) is an expression for only one interloper, including all types of interlopers in the survey. An expression for the interloper fraction for one line relative to the total number of galaxies including all interlopers is 
\begin{equation}
{f}_{i} = \frac{{N}_{Int,i}({z}_{Int,i})}{{N}_{SELG}({z}_{SELG})+\sum _{i}^{}{{N}_{Int,i}({z}_{Int,i})}}
\end{equation}
where the sum is over all interloping lines, and the total interloping fraction including interlopers from all interloping lines is
\begin{equation}
{f}_{tot} = \sum _{i}^{}{{f}_{i}}.
\end{equation}
The expression in Eq. (1) can be applied directly to a single interloping line for examining performance of secondary line identification and photometry.

An interloper's effect on the matter power spectrum is given in \cite{2015arXiv150705092P}) as: 
\begin{align}
{P}_{Int}(f\mid k,\mu,{z}_{SELG})={(1+{f}_{tot})}^{2}{P}_{SELG}(k,\mu ,{z}_{SELG})\nonumber\\
+\sum _{i}^{}{{{f}_{i}}^{2}{{\gamma}_{\perp}}^{2}{\gamma}_{\|}{P}_{Int,i}[q(k,\mu ),{\mu}_{q}(\mu),{z}_{Int,i}]} ,
\end{align}
where $\bold{x}$ represents the observed position assuming the interloping galaxy has a redshift ${z}_{SELG}$, and $\bold{y}$ is the real position of the galaxy. Note that we alter the expression to include the effect of multiple interloping lines, and ${\gamma}_{\|}$ and ${\gamma}_{\perp}$ are defined as, 
\begin{eqnarray}
({\bold x}_{\perp},{\bold x}_{\|})&=&({\gamma}_{\perp}{\bold y}_{\perp},{\gamma}_{\|}{\bold y}_{\|})\nonumber\\
{\gamma}_{\perp}&=&\frac{D({z}_{SELG})}{D({z}_{Int})}\nonumber\\
{\gamma}_{\|}& =&\frac{{\lambda}_{Int}H({z}_{Int})}{{\lambda}_{SEL}H({z}_{SELG})}.
\end{eqnarray}

A more comprehensive discussion along with examples of power spectrum and growth rate measurement biases due to interlopers can be found in \cite{2015arXiv150705092P}.
\section{PROGRAM DETAIL}

\subsection{COSMOS Mock Catalog}

The COSMOS Mock Catalog (CMC) \citep{2009ApJ...690.1236I, 2011A&A...532A..25J} predicts the emission line strengths of 564,555 high-redshift galaxies. Data read in the program can be categorised into three types: galaxy properties, band properties, and line properties. Galaxy properties are information inherent to the galaxies such as redshift, half-light radius, and distance modulus. Band properties includes the magnitude,magnitude depth and noise of a band, and line properties includes flux, noise and interloper fraction of a line, which are read into two different classes and passed around in the program. 

\subsection{CALIBRATION OF LUMINOSITY FUNCTION}

As luminosity functions are updated over time, it is necessary to update corresponding emission line fluxes in the mock catalog. Recall the Press-Schechter form of the luminosity function,
\begin{equation}
n(x;{\phi}^{*},\alpha)dx = {\phi}^{*}{x}^{\alpha}{e}^{-x}dx, \quad where \quad x = \frac{L}{L*},
\end{equation}
Since the luminosity function is not monotonic, it is difficult to obtain the calibrated luminosity by inverting the new luminosity function, so the cumulative luminosity function is used instead of the luminosity function itself,
\begin{equation}
n(>x;{\phi}^{*},\alpha) = \int_{x}^{\infty}{\phi}^{*}{x}^{\alpha}{e}^{-x}dx = {\phi}^{*}\Gamma (x;\alpha +1)
\end{equation}
To obtain the correct flux for a galaxy's emission line, the original cumulative luminosity function value needs to be calculated first, which gives the number density of galaxies with a luminosity greater than the given luminosity value. The calibrated luminosity is then computed by finding the inverse of the new cumulative luminosity function from the same number density. In other words,
\begin{equation}
{n}_{new} = {n}_{old}
\end{equation}
\begin{equation}
{L}_{new} ={n}^{-1}({n}_{new}; {\phi}^{*}{x}^{\alpha})
\end{equation}
Note that $n$ without any subscript is the old luminosity function.

\begin{table*}
\caption{The luminosity function for $H\alpha$ from \cite{2010MNRAS.402.1330G} is the luminosity function used in the CMC, and the luminosity functions for $H\alpha$ and OIII from \cite{0004-637X-779-1-34}, the luminosity functions used for calibrating the $H\alpha$ line and OIII doublet. For redshifts outside the range of specified luminosity function parameters, the parameters are set at the value of the closest boundary, instead of extrapolation. }
\begin{center}
\begin{tabular}{|c|c|c|c|}
\hline
source &line &redshift range&luminosity function parameters\\ \hline
\multirow{2}{10em}{Geach et al 2010} &
\multirow{2}{1em}{$H\alpha$} &
z $<$ 1.3 & ${\Phi}^{*} = 1.37\times{10}^{-3}$, $\alpha = -1.35 $,${L}^{*} = 5.1{(1+z)}^{3.1}\times{10}^{41}$\\&&
z $\geq$1.3 & ${\Phi}^{*} = 1.37\times {10}^{-3}, \alpha = -1.35, {L}^{*} = 6.8\times {10}^{42}$ \\ 
\hline
\multirow{2}{10em}{Colbert et al 2013} &
\multirow{2}{1em}{$H\alpha$} &
$0.3\leq z<0.9$ &${\Phi}^{*} = {10}^{-2.51}, \alpha = -1.27 ,{L}^{*} = {10}^{41.72}$\\&&
$0.9\leq z\geq 1.5$&${\Phi}^{*} = {10}^{-2.7}, \alpha = -1.43, {L}^{*} = {10}^{42.18}$\\ 
\hline
\multirow{2}{10em}{Colbert et al 2013} &
\multirow{2}{1em}{OIII}&
$0.7\leq z<1.5$&${\Phi}^{*} = {10}^{-3.28}, \alpha = -1.5 ,{L}^{*} = {10}^{42.39}$\\&&
$1.5\leq z\geq 2.3$&${\Phi}^{*} = {10}^{-3.60}, \alpha = -1.5, {L}^{*} = {10}^{42.83}$\\ 
\hline
\end{tabular}
\end{center}
\end{table*}%

We perform this procedure to recalibrate H$\alpha$ and OIII emission lines in the CMC.Table 1 shows the luminosity functions used in the CMC and the default luminosity functions used here for calibration. Users can provide their own luminosity functions for calibration when needed. To specify a user-defined luminosity function, parameters $({\Phi}^{*} , {L}^{*}, \alpha)$ must be provided for each redshift bin.

\subsection{PERTURBATION IN FLUX AND PHOTOMETRY} 

Obtaining the noise for emission lines is straightforward: given a sensitivity file containing the emission line noise as a function of observed wavelength, the program interpolates the noise for a particular line. If half-light radii are also included in the sensitivity file, a 2 dimensional interpolation can be performed.

We perturb photometric band flux according to the photometric error. In order to obtain the photometric error for a galaxy with finite radius, we use the magnitude depth given by:

\begin{equation}
{m}_{dep}={m}_{dep,0}-1.25{log}_{10}(1+({\frac{r}{r}_{0}})^{2}),
\end{equation}
Where ${m}_{dep,0}$ is the magnitude depth for a point source, and ${r}_{0}$ is the characteristic radial size of the point-spread function for that band. 

The program supports perturbing the spectral lines' flux and magnitude in photometric bands, which can give an estimation of the interloper fraction uncertainty, given by the standard deviation of the resulting interloper fractions for each simulation. The perturbation in flux is straightforward; for a spectral line with noise lower then a cut, the flux will be perturbed. We perturb the flux of the line using normal distribution, 
\begin{equation}
{f}_{new} = {f}_{original} + noise*\mathcal{N},
\end{equation}

Since data in photometric bands is given in magnitudes, it must be transformed to a flux before being compared to a threshold:

\begin{equation}
f = {10}^{-0.4m+48.6} erg {cm}^{-2} {s}^{-1}{Hz}^{-1}
\end{equation}

If the flux is higher than a threshold, then that band's flux will be perturbed in the same way as the flux for emission lines. \newline\newline

\subsection{IDENTIFYING INTERLOPERS}

Following the scheme in \cite{2015arXiv150705092P}, we use two methods are used in the program to distinguish interlopers from survey galaxies--secondary line identification and photometric cuts. From Eq. (2), we know the ratio between two emission lines wavelengths remains the same regardless of redshift. Thus, we can use a pair of lines to rule out the redshift ambiguity of the source of an emission line. Whenever we observe a line, we can try to find another line in the spectrum with a certain separation, and if the secondary line is found, then we can confirm that galaxy is a survey emission galaxy.

For interlopers that survived the secondary line identification procedure, colour cuts are applied to distinguish them from SELGs, note that the definition of colour here is referring to the difference in different photometric bands. Since the interloping galaxy in a sufficiently small redshift range will have a redshift different from the SELGs , the two may have different colours. This can be used to identify interlopers. 

After applying those cuts to the galaxy set, the leftovers are potential interlopers. Together with Eq. (5), the interloper fraction as a function of ${z}_{SELG}$ is outputted as an expected result from the program.

\section{RESULT}
	
To examine the power of cuts, we should use the expression from Eq. (1):

\begin{equation}
{f}_{I,cut}=\frac{{N}_{Int,i}({z}_{Int,i})}{{N}_{SELG}({z}_{SELG})+{{N}_{Int,i}({z}_{Int,i})}}
\end{equation}
	The expression in Eq. (15) contains only information about a specific interloping line and the SEL, and is therefore independent of other interloping lines. This can be used to investigate effect of cuts applied to that line.  To understanding the contribution of a particular line to the total interloper fraction among the whole survey, we can make use of Eq.(4). In Figs. 1 and 2, the left columns show the effect of cuts and the right columns show the contribution to total interloping fraction. 
	
	Since the denominator in Eq.4 depends on the interloper fraction of other interloping lines at the redshift bin, if the cutting conditions applied to other interloping lines are more effective than the interloping line being inspected at some particular redshift, the contribution to the total interloper fraction for that particular interloping line will be raised. That is the reason causing the total contribution of OII line in both surveys after applying secondary line cuts are higher than only SNR cuts in some redshift bin. The sum of total interloper fraction overall redshift should not be affected by this  .  \newline
	
	In figure 3, the total interloper fraction of the surveys are shown, which gives an estimation to the total contamination of the survey.

\begin{figure*}[htbp]
\includegraphics[width=\textwidth]{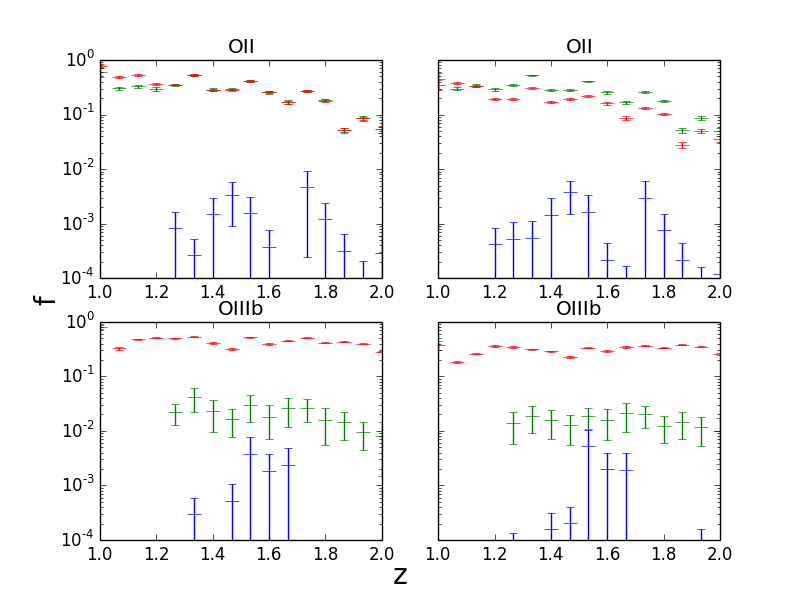}
\caption{Histograms of interloper fraction for Ha survey from OII line and OIIIb line.The interloper fraction here is the interloper fraction including the interloping line and the SEL only. The binning of these histograms is $\delta z$=0.1. The red line represents the interloper fraction for that line with SNR$>$7 only. The green line represents the interloper fraction with SNR$>$7 and secondary line identification, and the blue line represents the interloper fraction with SNR, secondary line identification, and photometric cuts. The right column is interloper fraction for that line in the whole survey, which corresponds to (4). And the left column is the interloping line and the SEL. Error comes from perturbing the flux noise in the simulation. Please note that the [OIIIb] line corresponds to [OIII 5007].}
\end{figure*}

\begin{figure*}[htbp]
\includegraphics[width=\textwidth]{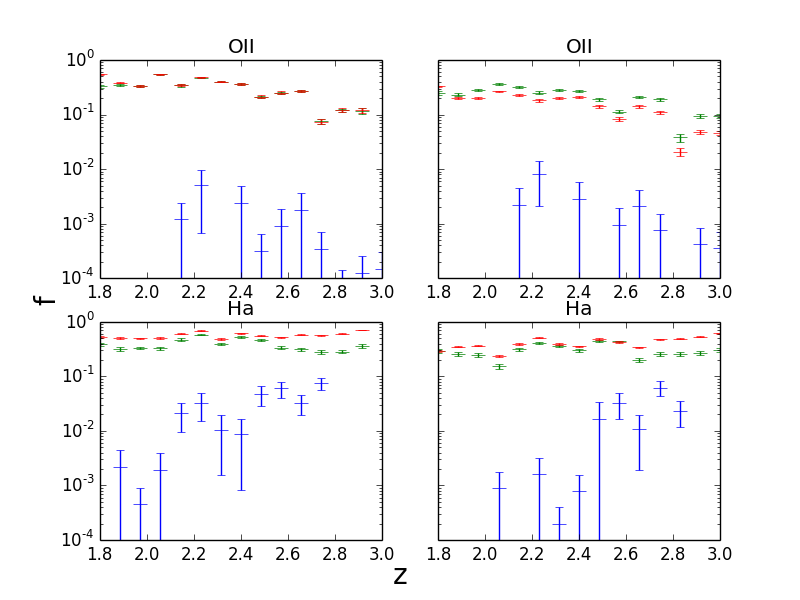}
\caption{Histograms of interloper fraction for OIIIb survey from OII line and Ha line, the configuration here is similar to plots in Figure1.}

\end{figure*}

\begin{figure*}[htbp]
\includegraphics[width=\textwidth]{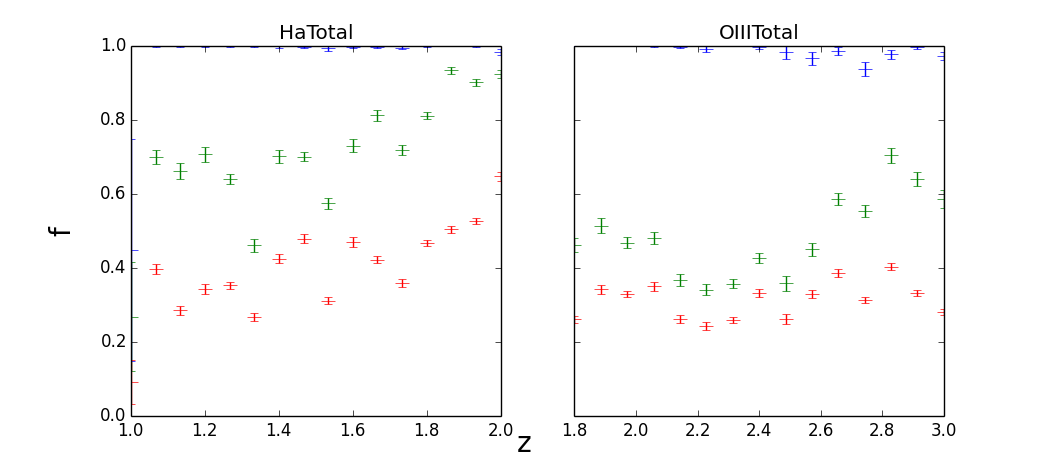}
\caption{The two plots here showing the total interloping fraction in the two survey, the configuration follows previous plots. }
\end{figure*}

\newpage

\section{CONCLUSION}

In this paper, we presented a code that estimates interloper fractions based on mock catalogues. The results obtained from this code agree with \cite{2015arXiv150705092P} . For a given survey sensitivity, the interloper fraction can be determined as a function of redshift. The part that identifies interlopers according flux and noise in emission lines and photometric bands can work on its own once the flux and noise are well defined in the program. Noise in photometric bands is now estimated based on the radial size and redshift of the object; other parameters can be added if needed. In order to modify the flux of emission lines according to user-defined functions, a separate routine can be used. Such calibration of luminosity functions may need end-user's further implementation, and there are examples given in a separate routine on Github. Using a catalog other than the CMC may require some modifications in how the program reads catalog data.

\section{ACKNOWLEDGEMENT}
K.W. and S.H were supported by NASA grant NASA 12-EUCLID11-0004 and 15-WFIRST15-0008 for this work. A.P. was supported by the McWilliams Fellowship of the Bruce and Astrid McWilliams Center for Cosmology.  

\bibliographystyle{apj}

\end{document}